\documentclass{article}
\usepackage{ismir,amsmath,cite,url}
\usepackage{graphicx}
\usepackage{color}
\usepackage{tabulary}
\usepackage{colortbl}
\usepackage{makecell}

\title{A Multimodal Approach Towards Emotion Recognition of Music Using Audio And Lyrical Content}

\twoauthors
  {Aniruddha Bhattacharya} {Senior Undergraduate,NIT Warangal \\ {\tt baniruddha2@student.nitw.ac.in}}
  {Kadambari K.V} {Assistant Professor,NIT Warangal \\ {\tt kadambari@faculty.nitw.ac.in}}

\begin{document}

\maketitle
\begin{abstract}
We propose \textit{MoodNet}-A Deep Convolutional Neural Network based architecture to
effectively predict the emotion associated with a piece of music given its
audio and lyrical content.We evaluate different architectures consisting of
varying number of two-dimensional convolutional and subsampling layers,followed
by dense layers.We use Mel-Spectrograms to represent the audio
content and word embeddings-specifically 100 dimensional word vectors, to
represent the textual content represented by the lyrics.We feed input data from
both modalities to our MoodNet architecture.The output from both the modalities
are then fused as a fully connected layer and softmax classfier is used to
predict the category of emotion.Using F1-score as our metric,our results show
excellent performance of MoodNet over the two datasets we experimented on-The
MIREX Multimodal dataset and the Million Song Dataset.Our experiments reflect
the hypothesis that more complex models perform better with more training
data.We
also observe that lyrics outperform audio as a better expressed modality and
conclude
that combining and using features from multiple modalities for prediction tasks
result in superior performance in comparison to using a single modality as
input.
\end{abstract}
\section{Introduction}\label{sec:introduction}
With the ever-increasing  amount of digital music online,there arises a need of effective organisation and retrieval of such amount of data.Although traditionally,the most common search and retrieval categories like artist and genre have received greater attention in music information retrieval(MIR) research,emotion based retrieval techniques too have been proven to be an effective criterion for MIR\cite{Author:01} receiving greater attention\cite{Author:02,Author:03}.

Generally,Music Emotion Recognition(MER) has relied on audio features like Mel-frequency cepstral coefficients (MFCCs) or mid-level features like chord\cite{5},rhythmic patterns\cite{5} etc for emotion recognition.

Lyrical features being semantically rich have also been widely used for emotion recognition,as their meanings convey emotions more clearly and are composed in accordance to the music signals.

For lyrical content analysis,mostly statistical natural language processing(NLP) techniques like bag of words\cite{10} and probabilistic latent semantic models\cite{11} have been used to extract textual features.

With recent evolution of powerful computational hardware like GPUs,Deep Neural Networks have been used successfully in audio content analysis and retrieval,with exceptional accuracy in tasks like speech recognition\cite{12} and computer vision\cite{13}

In computer vision, deep convolutional neural networks
(Deep CNNs) simulate
the behaviour of the human vision system and learn hierarchical
features, allowing object local invariance and robustness
to translation and distortion in the model\cite{14}.
They have shown state-of-the-art performance in
speech recognition\cite{12} and music related tasks like music segmentation\cite{16}.

Likewise,vector space representations have also proved to be am effective method of representing words\cite{20,21,22}.Using such representations along with Deep CNNs have shown great results in various NLP tasks like sentence classification\cite{17} and sentiment analysis\cite{18,19}.

In this paper,we propose we propose \textit{\textbf{MoodNet}}--a deep convolutional neural network based architecture,that combines the features obtained from both audio and lyrical modalities for classification of music based on mood.We use mel-spectrograms as input for the audio modality.For the corresponding lyrical content,we use the vector representation of words present in the lyrical content as input to the network.Both of these representations are used as inputs to the Deep CNNs which output a single feature vector(for each modality).The two vectors are combined and are classified using a softmax classifier.

We disuss CNNs for audio content and textual content analysis in Section 2 and Section 3 respectively.The \textit{MoodNet} architecture is introduced in Section 4 followed by experiments and results in Section 5. We end our paper with a conclusion and the scope for future work in Section 6.

\section{CNNs in Audio content analysis}

\subsection{Motivation}

CNNs are motivated by our perception of vision where neurons capture local information and higher level information is obtained\cite{14}.CNNs
are therefore designed to provide a way of learning robust
features that respond to certain visual objects with translational and distortion invariance.These advantages
often work well with audio signals too.Deep CNNs learn features hierarchically,learning lower level features at the shallow end and hierarchically learning complex and higher-level features at the deeper ends\cite{14}.

Audio analysis tasks uses CNNs with the underlying assumption that auditory events are detectable by observing their time-frequency representation.As emotion of a song represents a high-level feature as compared to beat,chords,tempo as mid-level features,this hierarchical nature aligns with the motivations behind the architecture of Deep CNNs.

\subsection{Representation}

Mel-spectrograms have been one of the most widespread features used
for various audio analysis tasks like music auto-tagging and latent feature learning . The use of the
mel-scale is supported by domain knowledge about the human
auditory system\cite{25} and has been empirically proven
by impressive performance gains in various tasks\cite{26}.

\begin{figure}
 \centerline{\framebox{
 \includegraphics[width=\columnwidth]{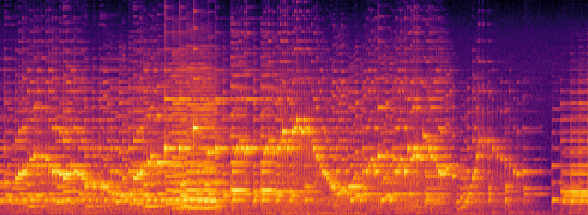}}}
 \caption{Mel-spectrogram representation of audio.Time and frequency are represented by Y and X-axes respectively.}
 \label{fig:example}
\end{figure}

The visual representation of audio mel-spectrogram is used as input to the MoodNet architecture(\figref{fig:example}).

\subsection{Convolutions}

A convolution layer of size l×b×d learns d features
of l×b, where l refers to the height and b refers to the width of the learned kernel. The kernel size also represents the maximum span of a component it can capture. If the kernel size is too small, the corresponding convolutional layer
would fail to learn a proper representation of the distribution of the data. For this reason, relatively larger dimensional
kernels are often preferred\cite{27}.

The convolution axes are an important aspect of
convolution layers. 2D convolutions generally perform better than 1D convolutions as the former can learn both temporal and spectral structures and have been used in tasks like boundary detection\cite{16} and chord recognition\cite{27}

\subsection{Pooling}

The pooling operation results in reduction of the feature map size with an operation,
usually a max function. It is widely used in most of the modern CNN architectures.Pooling applies subsampling to reduce the size of feature
map.While doing so ,instead of preserving information about the whole input,it only
tries to preserve the information of an activation in
the region.
The non-linear behaviour of subsampling also provides
distortion and translation invariances. For smaller pooling sizes
,the network cannot have enough distortion
invariance.On the other hand,if it is too large, many feature locations may be left out when needed. Normally, the pooling axes should
match the convolution axes.

\section{CNNs in Lyrical content analysis}
\subsection{Motivation}
\subsubsection{Word Embeddings}
Semantic representation of words is a challenging task in natural
language processing.With the recent development
of neural word representations models\cite{29,30,31},word embeddings have provided a broad scope
for distributional semantic models. For the first time, distributed
representations of words makes it possible to capture semantics of words; including even the shift in meaning of words over
time\cite{32}. Such capability explains the recent successful
switch in the field of natural language processing from linear models over sparse inputs,
e.g. support vectors machines(SVMs) and logistic regression,
to non-linear neural-network models over dense inputs.
As a result,models that rely on word embedding have
been very successful in recent years, across a large spectrum of language processing tasks\cite{33}.
Word embeddings based on neural networks are prediction-based models.For a network to  learn  distributed representations
for words,it learns its parameters by
predicting the correct word (or its context) in a suitable text window
over the training corpus.
While the main objective of training the entire network is to learn superior
parameters, word vector representations are based upon the idea
that similar words are closer together. In linguistics, this is
known as “Distributional Hypothesis”. This very idea
is beneficial for extracting features from text represented in
a 'natural' way; especially for understanding the context of word
use in mood prediction. Since this notion is viable for any
natural language, we take advantage of that and apply it
to musical lyrics.
\begin{figure}
 \centerline{
 \includegraphics[width=\columnwidth]{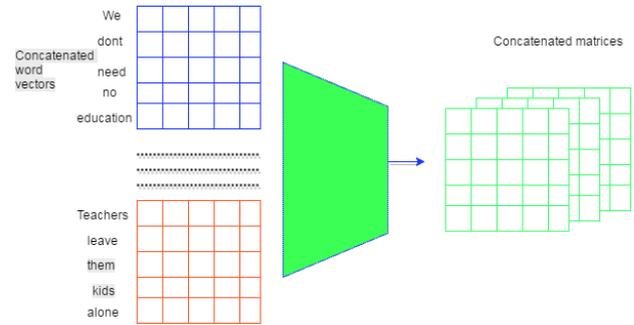}}
 \caption{The construction of the text input from lyrical content as a 3 dimensional matrix.}
 \label{harm}
\end{figure}
\subsection{Representation}

Each word in the lyrics of a song is a vector.So all the words in a sentence are vectors which ,when concatenated represent a two-dimensional matrix.Similarly,multiple lines of sentences when concatenated represent a three-dimensional matrix.

The resultant input is similar to that of an image with multiple channels and thus serves as input to our Deep CNN architecture.The representation is shown in \figref{harm}.

\section{MoodNet Architecture}

Figure 2 shows one of the proposed architectures,
a 4-layer MoodNet architecture which consists of 4 convolutional
layers and 4 max-pooling layers.For the audio content , this network takes
a log-amplitude mel-spectrogram sized 96×1366 as input.
\definecolor{Gray}{gray}{0.85}
\definecolor{LightGray}{gray}{0.9}
\definecolor{LightestGray}{gray}{0.75}
\begin{table}[h]
\begin{center}
\begin{tabular}{ |c|c|c| }
\hline
MoodNet-3 & *MoodNet-4 & **MoodNet-5 \\
 \hline
 \hline
\rowcolor{Gray}

 \multicolumn{3}{| c |}{
Mel-spectrogram \textit{(input: 96$\times$1366$\times$1)}}\\
 \hline
 \hline
 \multicolumn{3}{| c |}{ \makecell{Conv $3$$\times$$3$$\times$$128$}} \\
 \hline
  \multicolumn{3}{| c |}{ \makecell{MP ($2$, $4$) \textit{(output: 48$\times$341$\times$128)}} }\\
 \hline
  \multicolumn{3}{| c |}{ \makecell{Conv $3$$\times$$3$$\times$$256$}}\\
 \hline
  \multicolumn{3}{| c |}{ \makecell{MP ($2$, $4$) \textit{(output: 24$\times$85$\times$256)}} }\\
 \hline
  \multicolumn{3}{| c |}{ \makecell{Conv $3$$\times$$3$$\times$$512$}}\\
 \hline
  \multicolumn{3}{| c |}{ \makecell{MP ($2$, $4$) \textit{(output: 12$\times$21$\times$512)}}  }\\
 \hline
 \rowcolor{LightGray}
  \multicolumn{3}{| c |}{ \makecell{*Conv $3$$\times$$3$$\times$$1024$} }\\
 \hline
 \rowcolor{LightGray}
  \multicolumn{3}{| c |}{ \makecell{*MP ($3$, $5$) \textit{(output: 4$\times$4$\times$1024)}} }\\
 \hline
 \rowcolor{LightestGray}
  \multicolumn{3}{| c |}{
 \makecell{**Conv $3$$\times$$3$$\times$$2048$}}\\
 \hline
 \rowcolor{LightestGray}
   \multicolumn{3}{| c |}{ \makecell{**MP ($4$, $4$) \textit{(output: 1$\times$1$\times$2048)}} }\\
 \hline 

 \hline
 \hline
\rowcolor{Gray}
 \multicolumn{3}{| c |}{
  Flatten 2048$\times$1}\\
 \hline
 \end{tabular}
\caption{The configurations of 3, 4, and 5-layer architectures for the \textbf{audio} modality.The darker layers show the additional layers for 4 and 5-layer architectures}
\label{table1}
\end{center}
\end{table}

\definecolor{Gray}{gray}{0.85}
\definecolor{LightGray}{gray}{0.9}
\definecolor{LightestGray}{gray}{0.75}
\begin{table}[h]
\begin{center}
\begin{tabular}{ |c|c|c| }
\hline
MoodNet-3 & *MoodNet-4 & **MoodNet-5 \\
 \hline
 \hline
\rowcolor{Gray}

 \multicolumn{3}{| c |}{
Mel-spectrogram \textit{(input: 100$\times$10$\times$20)}}\\
 \hline
 \hline
 \multicolumn{3}{| c |}{ \makecell{Conv $3$$\times$$3$$\times$$6$}} \\
 \hline
  \multicolumn{3}{| c |}{ \makecell{MP ($2$, $2$) \textit{(output: 49$\times$5$\times$120)}} }\\
 \hline
  \multicolumn{3}{| c |}{ \makecell{Conv $3$$\times$$3$$\times$$256$}}\\
 \hline
  \multicolumn{3}{| c |}{ \makecell{MP ($2$, $2$) \textit{(output: 24$\times$4$\times$256)}} }\\
 \hline
  \multicolumn{3}{| c |}{ \makecell{Conv $3$$\times$$3$$\times$$512$}}\\
 \hline
  \multicolumn{3}{| c |}{ \makecell{MP ($2$, $2$) \textit{(output: 12$\times$2$\times$512)}}  }\\
 \hline
 \rowcolor{LightGray}
  \multicolumn{3}{| c |}{ \makecell{*Conv $3$$\times$$3$$\times$$1024$} }\\
 \hline
 \rowcolor{LightGray}
  \multicolumn{3}{| c |}{ \makecell{*MP ($3$, $2$) \textit{(output: 4$\times$1$\times$1024)}} }\\
 \hline
 \rowcolor{LightestGray}
  \multicolumn{3}{| c |}{
 \makecell{**Conv $3$$\times$$3$$\times$$2048$}}\\
 \hline
 \rowcolor{LightestGray}
   \multicolumn{3}{| c |}{ \makecell{**MP ($4$, $2$) \textit{(output: 1$\times$1$\times$2048)}} }\\
 \hline 

 \hline
 \hline
\rowcolor{Gray}
 \multicolumn{3}{| c |}{
  Flatten 2048$\times$1}\\
 \hline
 \end{tabular}
 \caption{The configurations of 3, 4, and 5-layer architectures for the \textbf{text} modality.The darker layers show the additional layers for 4 and 5-layer architectures.Note that zero padding has been used whenever required to avoid dimensionality reaching zero}
\label{table10}
\end{center}
\end{table}

Similarly,for the lyrical content,the entire corpus is first searched for the line with maximum length,This shall be the width of the three-dimensional matrix input.For all other sentence,the edges are padded with zeroes to match the maximum-width,resulting in a uniform width three-dimensional matrix.

Each word is represented as a vector of length 100.We use the publicly available GloVe\cite{34} vectors, which were trained on a corpus of 6 Billion tokens from a 2014 Wikipedia dump. The vectors are of dimensionality 100, trained on the non-zero entries of a global word-word co-occurrence matrix, which tabulates how frequently words co-occur with one another in a given corpus.Thus,each word represents a vector,each sentence a matrix,hence each instance of multiple lined sentence i.e the lyrics of an entire song is represented as a three-dimensional matrix.

\begin{figure}
 \centerline{
 \includegraphics[width=\columnwidth]{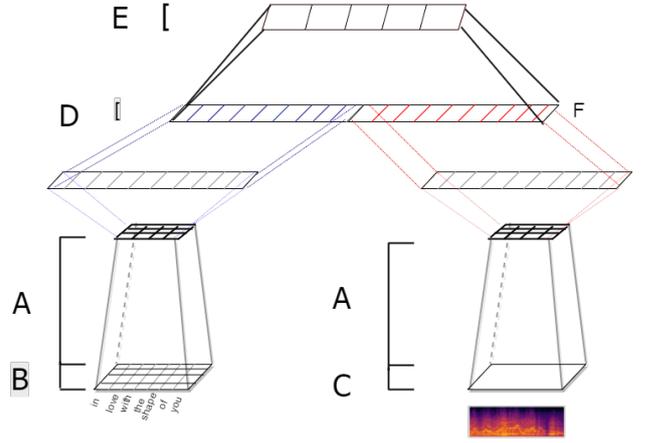}}
 \caption{Overall architecture of MoodNet.Here \textbf{A}--The Convolutional Layers;\textbf{B}--The text input as sentence;\textbf{C}--The Mel-Spectrogram input for audio;\textbf{D}--The Fully connected layer with successive dense layers and dropout;\textbf{E}--Softmax output;\textbf{F}--Modalities combined}
 \label{figmain}
\end{figure}

Both the lyrical and audio components are fed into our Deep CNN architecture,which results in two feature vectors for each modality.As referred in \tabref{table1} and \tabref{table10},we obtain 2048 units from each modality after flattening the final layer from the Deep CNN architecture.We now combine the modalities,resulting in a 1D vector of 5096 units.We now progressively add Dense layers and a 20 \% dropout rate and successively reduce the size of the layer from 5096 to 2048 units,1024 units,512 units and finally 256 units,using ReLU activation.Finally,we obtain a fully connected layer of size 5.This layer is now classified using a softmax classifier.The overall architecture is represented in \figref{figmain}.

The architecture is extended to deeper ones with 4 and 5 layers.The number of feature maps
and subsampling sizes for the audio modality are summarised in \tabref{table1}.Similarly,the deep architecture for text modality is summarized in \tabref{table10}.
\section{Experiments and Results}

\subsection{Overview}
We used two datasets to evaluate our MoodNet architecture,
the MIREX Multimodal dataset\cite{34}(Dataset I) and the Million Song
Dataset\cite{35}(Dataset II).

We test three architectures (MoodNet-3,4,5) in both Dataset I and Dataset II.
In both datasets, the audio was trimmed as 29.0 clips (the shortest signal
in the dataset) and  downsampled to 12 kHz. The
hop size was set to 256 samples (21 ms) during time-frequency
transformation, resulting in an output of 1,366 frames in total.

For the lyrics, Dataset I  already contains the lyrics.For Dataset II ,we selected a subset of the song names and scraped their corresponding lyrics from \textit{lyrics.wikia.com} if they were available,else removed them from the dataset.

As each word represents a 100 dimensional vector,each sentence in the lyrics represented a matrix(by concatenating the vectors vertically).Again,a number of such sentences,when concatenated,would represent a 3-dimensional matrix.Of course,we take necessary steps to handle problems like variable length of sentences for a song,or variable number of lines for different songs.

The audio and textual components are used as inputs to the MoodNet architecture.

We used ADAM adaptive optimisation \cite{36} on Keras
\cite{37} and Theano \cite{38} framework during the experiments.Categorical 
cross-entropy function has been used since as it results in faster
convergence than distance-based
functions such as mean squared error and mean absolute
error.
\subsection{Dataset I:MIREX Multimodal}
\begin{figure}[h]
 \centerline{
 \includegraphics[width=\columnwidth]{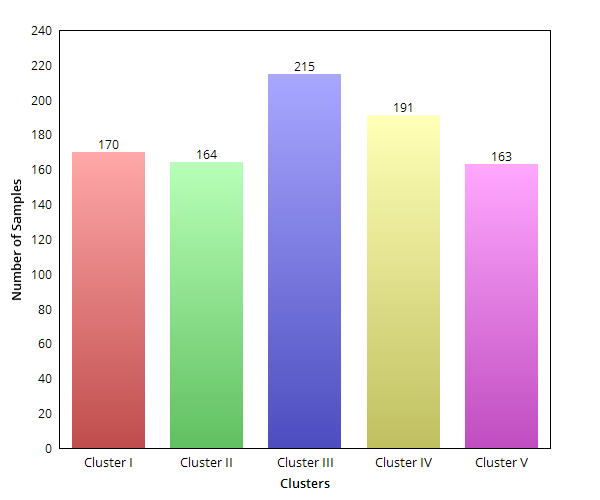}}
 \caption{The distribution of samples among the five clusters in the MIREX Multimodal Dataset that we obtained.}
 \label{volee}
\end{figure}

 The MIREX Multimodal dataset has a total of 903 30-second clips, each of which belongs to one of the five clusters (as shown in \tabref{tab2}). Each cluster contains different numbers of clips, say, 170 clips in cluster I, 164 clips in cluster II, 215 clips in cluster III, 191 clips in cluster IV, and 163 clips in cluster V.The distribution has been represented in \figref{volee}.
 
\begin{table}[h]
 \begin{center}
 \begin{tabular}{|l|l|}
 \hline
 Cluster & Mood \\
  \hline
   \hline
  I & passionate, rousing, confident, boisterous \\
  \hline
  II  &  cheerful, fun, sweet,amiable \\
  \hline
  III & poignant, wistful, bittersweet, autumnal \\
  \hline
  IV &  humorous, silly, campy, quirky, witty \\
  \hline
  V & aggressive, fiery, intense, volatile\\
  \hline
 \end{tabular}
\end{center}
 \caption{Emotion Categories and their defined clusters used in MIREX Multimodal Dataset}
 \label{tab2}
\end{table}

We used F1 score as the accuracy metric for our experiment,as it considers both the precision 'p' and the recall 'r' value to compute the score.
The results obtained by our architecture are summarized in \tabref{tab3}.

\begin{table}[h]
 \begin{center}
 \begin{tabular}{|l|l|}
 \hline
 Architecture & F-measure \\
  \hline
   \hline
  MoodNet-3 & 72.3\\
  \hline
  MoodNet-4 & \textbf{76.34}\\
  \hline
  MoodNet-5 & 75.68\\
  \hline
 \end{tabular}
\end{center}
 \caption{F-score obtained by various MoodNet architectures on the MIREX Multimodal dataset}
 \label{tab3}
\end{table}

\subsection{Dataset II:Million Song Dataset}

We also evaluated our MoodNet architecture using the Million
Song Dataset (MSD) with last.fm tags. We select the
top 50 tags and extracted the mood based tags only.Among the variety of tags present,we clustered the tags according to \tabref{tab2}.We selected  subset of 50,000 samples from MSD and searched for the corresponding lyrics from \textit{lyrics.wikia}.
\begin{figure}[h!]
 \centerline{
 \includegraphics[width=\columnwidth]{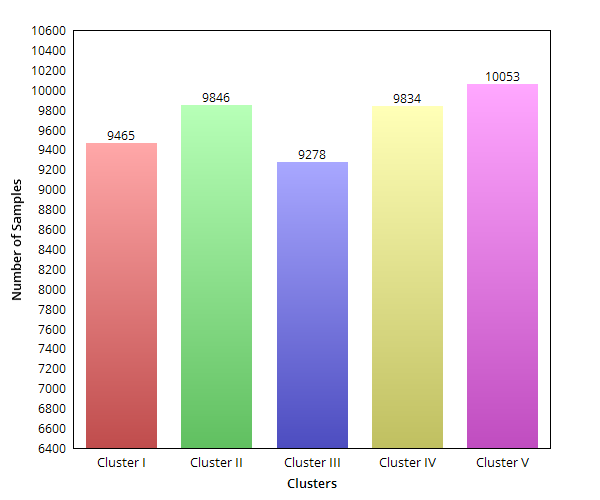}}
 \caption{The distribution of samples among the five clusters in the subset of the Million Song Dataset that we obtained.}
 \label{chut23}
\end{figure}
We removed the songs whose lyrics were not found.Our labels for each song was made by clustering the obtained tags and grouping them into one of the five clusters as described in \tabref{tab2}.Thus,the total dataset was reduced to 48,476
(40,476 for training and 8000 for validation).The distributionof samples across the datset has been represented by \figref{chut23}.

We followed the same process of representing audio as mel-spectrograms and lyrics as word embeddings as in Dataset I.

We used F-1 score as the accuracy metric in our experiment with MSD.The results obtained are summarized in \tabref{tab4}.
\begin{table}[h]
 \begin{center}
 \begin{tabular}{|l|l|}
 \hline
 Architecture & F-measure \\
  \hline
   \hline
  MoodNet-3 & 66.28\\
  \hline
  MoodNet-4 & 69.73\\
  \hline
  MoodNet-5 & \textbf{71.29}\\
  \hline
 \end{tabular}
\end{center}
 \caption{F-score obtained by various MoodNet architectures on the Million Song Dataset}
 \label{tab4}
\end{table}

\subsection{Modalities and Performance}

We also experimented with different modalities as inputs.We supplied only audio as input ,only text(lyrical content) and both audio and lyrical modalities.

From the results we obtained,as shown in \tabref{tab5},it is clear that when considering single modality as input,lyrics outperform audio.This result is expected as lyrics convey meaning and emotion more explicitly,than Mel-spectrograms of audio .It is also observed that when both audio and text modalities are combined ,they outperform the results obtained from a single modality input.This leads us to conclude that combining audio and text modalities leads to an increase in accuracy in emotion detection. 

\begin{table}[h!]
 \begin{center}
 \begin{tabular}{|l|l|l|l|}
 \hline
 Dataset & Audio & Lyrics & Audio+Lyrics \\
  \hline
   \hline
  MIREX Multimodal & 56.46 & 62.39 & \textbf{66.28}\\
  \hline
  MSD & 58.34 & 64.79 & \textbf{69.73}\\
  \hline
 \end{tabular}
\end{center}
 \caption{F-score obtained by MoodNet-4 architecture for audio and text modalities as inputs,both individually and combined}
 \label{tab5}
\end{table}

\section{Conclusion and Future Work}

We presented \textit{MoodNet} -an emotion detection model based on Deep Convolutional Neural Networks. It was shown
that our MoodNet architecture ,based on Deep Convolutional Neural Networks with 2D convolutions can be effectively
used for emotion detection.We tested our hypothesis on two datasets.
In both datasets,we tested our architectures with both audio and text input ,individually and combined.For audio,we used mel-spectrograms as input.For the text modality,we used word embeddings as input.

Our results show that lyrics as a single modality input outperforms audio.Also,combining both the modalities gave a better performance in both our datasets.Thus,we conclude that a combination of various modalities as input results in a better representation of a song as a whole.

It is to be noted that we didn't use Long Short-Tem Memory(LSTM) based Recurrent Neural Networks(RNNs) as our task involved detecting the emotion of an entire piece of music as a whole.In the future,we also plan to build an architecture capable of showing dynamic temporal behaviour.RNNs may be useful in that regard.

As future work,we also plan to explore video as an additional input modality.Also,mood based song recommender systems based on our architecture,should effectively help users discover new music and tackle the cold-start problem associated with collaborative filtering,used in most current recommender systems.

\bibliography{abs_cite}

\begin{thebibliography}{10}

\bibitem{38}
Frederic Bastien, Pascal Lamblin, Razvan~Pascanu and´ James~Bergstra, Ian
  Goodfellow, Arnaud Bergeron, Nicolas Bouchard, David Warde-Farley, and Yoshua
  Bengio.
\newblock Theano: new features and speed improvements.
\newblock {\em arXiv:1211.5590}, 2012.

\bibitem{35}
Thierry Bertin-Mahieux, Daniel~PW Ellis, Brian Whitman, and Paul Lamere.
\newblock The million song dataset.
\newblock {\em Proc. of the 12th International Society for Music Information
  Retrieval Conference,(ISMIR)}, pages 591–--596, 2011.

\bibitem{37}
Francois Chollet.
\newblock {\em Keras:Deep learning library for theano and tensorflow}.
\newblock 2015.

\bibitem{22}
R.~Collobert and J.~Weston.
\newblock A unified architecture for natural language processing: Deep neural
  networks with multitask learning.
\newblock {\em Proc. of the 25th international conference on Machine learning},
  pages 160--167.

\bibitem{26}
Sander Dieleman and Benjamin Schrauwen.
\newblock End-to-end learning for music audio.
\newblock {\em IEEE International Conference on Acoustics, Speech and Signal
  Processing (ICASSP), 2014}, pages 6964–--6968, 2014.

\bibitem{18}
C.~N. dos Santos and Gatti.M.
\newblock Deep convolutional neural networks for sentiment analysis of short
  texts.
\newblock {\em COLING-2014}, pages 69--78.

\bibitem{Author:03}
Lu.L et~al.
\newblock Automatic mood detection and tracking of music audio signals.
\newblock {\em IEEE Trans. Audio,Speech and Language Processing}, 14(1):5--18,
  2006.

\bibitem{Author:01}
A~Friberg.
\newblock In {\em Digital Audio Emotions – An Overview of Computer Analysis
  and Synthesis of Emotional Expression in Music}, pages 1--6, 2008.

\bibitem{11}
T~Hofmann~et.al.
\newblock Probabilistic latent semantic indexing.
\newblock {\em Proc. ACM SIGIR}, pages 50--57, 1999.

\bibitem{27}
Eric~J Humphrey and Juan~P Bello.
\newblock Rethinking automatic chord recognition with convolutional neural
  networks.
\newblock {\em 11th International Conference on Machine Learning and
  Applications}, 2:357--362, 2012.

\bibitem{36}
Diederik~P. Kingma and Jimmy Ba.
\newblock Adam: A method for stochastic optimization.
\newblock {\em CoRR}, 2014.

\bibitem{13}
A.~Krizhevsky, I.~Sutskever, and G.~Hinton.
\newblock Imagenet classification with deep convolutional neural networks.
\newblock {\em NIPS}, 2012.

\bibitem{31}
Q.~V. Le and T.~Mikolov.
\newblock Distributed representations of sentences and documents.
\newblock {\em Proc. of The 31st International Conference on Machine Learning},
  pages 1188--1196, 2014.

\bibitem{14}
Yann LeCun and Yoshua Bengio.
\newblock Convolutional networks for images, speech, and time series.
\newblock {\em The handbook of brain theory and neural networks}, page
  3361(10), 1995.

\bibitem{33}
T.~Luong, R.~Socher, and C.~D. Manning.
\newblock Better word representations with recursive neural networks for
  morphology.
\newblock {\em CoNLL}, pages 104--113, 2013.

\bibitem{32}
C.~D. Manning.
\newblock Computational linguistics and deep learning.
\newblock {\em COLING}, 41(4):701–--707, 2015.

\bibitem{30}
T.~Mikolov, K.~Chen, G.~Corrado, and J.~Dean.
\newblock Distributed representations of words and phrases and their
  compositionality.
\newblock {\em Proc. Advances in Neural Information Processing Systems},
  26:3111--–3119, 2013.

\bibitem{29}
T.~Mikolov, K.~Chen, G.~Corrado, and J.~Dean.
\newblock Efficient estimation of word representations in vector space.
\newblock {\em ICLR}, 2013.

\bibitem{25}
Brian~CJ Moore.
\newblock {\em An introduction to the psychology of hearing}.
\newblock Brill, 2012.

\bibitem{21}
F.~Morin and Y.~Bengio.
\newblock Hierarchical probabilistic neural network language model.
\newblock {\em Proc. of the international workshop on artificial intelligence
  and statistics}, pages 246--252, 2005.

\bibitem{34}
R.~Panda, R.~Malheiro, B.~Rocha, A.~Oliveira, and R.~P. Paiva.
\newblock Multi-modal music emotion recognition: A new dataset methodology and
  comparative analysis.
\newblock {\em Proc. CMMR}, pages 570--582, 2013.

\bibitem{5}
Lu~Qi, Xiaoou Chen, Deshun Yang, and Jun Wang.
\newblock Boosting for multi-modal music emotion.
\newblock {\em 11th International Society for Music Information and Retrieval
  Conference}, 2010.

\bibitem{12}
Tara~N Sainath, Abdel rahman Mohamed, Brian Kingsbury, and Bhuvana Ramabhadran.
\newblock Deep convolutional neural networks for lvcsr.
\newblock {\em IEEE International Conference on Acoustics, Speech and Signal
  Processing (ICASSP)}, pages 8614--8618, 2013.

\bibitem{10}
F~Sebastiani.
\newblock Machine learning in automated text categorization.
\newblock {\em ACM CSUR}, 34(1):1--47, 2002.

\bibitem{16}
Karen Ullrich, Jan Schluter, and Thomas Grill.
\newblock Bound- ¨ ary detection in music structure analysis using
  convolutional neural networks.
\newblock {\em Proc. of the 15th International Society for Music Information
  Retrieval Conference, ISMIR}, 2014.

\bibitem{Author:02}
Y.H. Yang~et al.
\newblock In {\em Toward multi-modal music emotion classification}, pages
  70--79, 2008.

\bibitem{20}
Y.Bengio, R.~Ducharme, P.~Vincent, and C.~Janvin.
\newblock A neural probabilistic language model.
\newblock {\em The Journal of Machine Learning Research}, 3:1137--1155, 2003.

\bibitem{17}
Kim Yoon.
\newblock Convolutional neural networks for sentence classification.
\newblock {\em Proc. of the 2014 Conference on Empirical Methods in Natural
  Language Processing (EMNLP)}, pages 1764--1751, 2014.

\bibitem{19}
Xiang Zhang, Junbo Zhao, and Yann LeCun.
\newblock Character-level convolutional networks for text classification.
\newblock {\em Advances in Neural Information Processing Systems}, pages
  649--657, 2015.

\end{thebibliography}

\end{document}